\documentclass{iopart}
\usepackage{iopams}  
\begin{document}
\title{Superfluidity and Stationary Space-Times}
\author{George Chapline$^{1}$, Pawel O. Mazur$^{2}$}
\address{ $^{1}$ Physics and Advanced Technologies Directorate, Lawrence
Livermore National Laboratory, Livermore, CA 94550, USA}
\ead{chapline1@llnl.gov}
\vskip .1cm
\address{$^{2}$ Department of Physics and Astronomy, University of South Carolina, \\ 
Columbia, SC 29208, USA}
\eads{\mailto{mazur@physics.sc.edu}, \mailto{mazurmeister@gmail.com}}
\vskip .1cm
\address{December 2008}

\begin{abstract}
A connection between superfluidity and gravitation is established for physical stationary gravitational fields. 
We show that the spinning cosmic string metric describes the gravitational field associated 
with the single vortex in a superfluid condensate model for space-time outside the vortex core. 
This metric differs significantly from the usual acoustic metric for the Onsager-Feynman vortex. 
We also consider the question of what happens when many vortices are present, and show that on large scales 
a G\"{o}del-like metric emerges. In both the single and multiple vortex cases the presence 
of closed time-like curves is attributed to the breakdown of superfluid rigidity.
\end{abstract}

\maketitle

The various developments of quantum field theory in curved space-time have left the false impression that general relativity
and quantum mechanics are compatible as long as one considers the length scales 
well above the Planck length $L=\sqrt{\frac{G\hbar}{c^{3}}}$. Indeed, viewing Gravitation and the Standard Model 
of elementary matter excitations, the so-called elementary particles, as an effective field theory (EFT) with a cutoff that is based on 
the Wilsonian renormalization group (RG) paradigm may lead to incorrect inferences about the microscopic nature of these phenomena. 
Certainly, it was historically an unfortunate development, that an unnecessary emphasis was placed on the high energy-momentum (UV) behavior of scattering amplitudes in `quantum Einstein gravity', which has blinded most of the early researchers \cite{Feyn63}, 
and their later followers to the subtleties of the global properties of the gravitational vacuum medium. 
We have in mind the infrared (IR) behavior of the gravitational and the Standard Model interactions 
of massless elementary excitations in the physically relevant case of the finite positive 
vacuum energy density $\epsilon_{_{vac}}={\mu}^{4}({\hbar c})^{-3}$. 
The physical gravitational vacuum state in this case is the de Sitter universe \cite{ASE31,PAMD1954,Bj03,M88,M92}.  
It was recognized that the physical gravitational vacuum state is a highly correlated quantum state 
of a new kind of matter which constituents were called gravitational atoms \cite{M88,M92,M95,M96,M97a,M97b,M97c,M98,M07}. 
In particular, in \cite{M88} a droplet model of fission of small black holes was proposed. 
The fundamental role played by quantum entanglement in the quantum state of a huge number 
of strongly interacting bosonic constituents of the gravitational vacuum medium in the explanation 
of the underlying microscopic mechanism responsible for the selection of very small values of the cosmological constant 
was strongly emphasized by one of the authors \cite{M92}.  
   
Incidentally, it was Einstein who first applied the reasoning based on symmetry and the power counting 
(the minimal number of derivatives of basic fields compatible with symmetries entering the action principle), 
standing behind the effective field theory (EFT) approach, when he had derived his field equations of relativistic gravitation. 
Using the arguments based on symmetry and power counting in the sense of EFT 
Einstein had introduced three macroscopic operationally defined constants $c$, $G$, and $\Lambda$ 
characterizing the physical gravitational vacuum. 
These three macroscopic constants are the velocity of light in vacuum $c$, 
which is at the same time identical to the speed of propagation of gravitational perturbations, 
the Newton-Cavendish gravitational constant $G$, and the cosmical or cosmological constant $\Lambda$. 
Eddington \cite{ASE31} and later Dirac \cite{PAMD1954} thought that the $\Lambda$ term 
in the Hilbert-Einstein action describes a very special kind of the de Sitter-Lorentz invariant medium. 
The unique combination of these three macroscopic constants characterizing the gravitational vacuum medium 
is the positive vacuum energy density $\epsilon_{_{vac}}$ and the negative vacuum pressure $P_{_{vac}}$, 
where $-P_{_{vac}}=\epsilon_{_{vac}}=\frac{c^{4}\Lambda}{8\pi G}$.  

It turns out that certain predictions of classical general relativity such as closed
time-like curves and event horizons are in conflict with a quantum
mechanical description of space-time itself. In particular, a
quantum mechanical description of any system requires a universal
time. In practice, universal time is defined by means of
synchronization of atomic clocks, but such synchronization is not
possible in space-times with event horizons or closed time-like
curves. It has been suggested \cite{M07,CM04,GC92} that the way a global
time is established in Nature is via the occurrence of
off-diagonal long-range quantum coherence in the vacuum state.
This leads to a very different picture of compact astrophysical
objects from that predicted by general relativity
\cite{M88,M92,M95,M96,M97a,M97b,M97c,M98,M07,M01,M04,CHLS00,CHLS03,RBL03}. 

We wish to point out the salient differences between the general relativistic description 
of rotating space-times and the picture offered by the assumption that the vacuum state 
is a quantum condensate of heavy bosons \cite{M88,M96,M97a,M97b,M97c,M98,M07,CM04,M01,M04}. 
In order to make the basic observations on the connection between superfluidity and gravitation explicit 
we shall consider a model of a \emph{weakly interacting} boson system in the Bogoliubov 
mean field approximation \cite{CM04}. This approximation is quite inadequate when it comes 
to the computation of the ground state energy density of the finite-size system 
of a large number of massive bosons. We have in mind the actual computation 
of the so-called cosmological constant $\Lambda$ \cite{M88,M07}.    

It has been recognized for a long time that general relativity fails to describe accurately 
the physical situation in the regions of extremely high tidal forces or curvature
singularities of the type of a Big Bang or the interior of an analytically continued 
black hole solutions. Generally, this breakdown of general relativity was
considered inconsequential because it was supposed to occur on
Planckian length scales. In this case a rather soothing philosophy
was adopted to the effect that some mysterious and still unknown
quantum theory of gravitation will take care of the difficulty by `smoothing out' the curvature singularities. 
It was recognized only recently that the physics of event horizons is a second
example of the breakdown of general relativity but this time on the
macroscopic length scales \cite{M88,M92,M96,M97a,M97b,M97c,M98,M07,CM04,M01,M04,CHLS00,CHLS03,RBL03}. 
In the following we consider a third kind of the breakdown of general
relativity on the macroscopic length scales, associated with the
occurrence of closed time-like curves (CTC). CTCs occur frequently
in analytically extended space-times described by general
relativity once there is rotation present in a physical system
under consideration, which is quite common in nature.

The well known example of a solution to the Einstein equations
where CTCs occur is the G\"{o}del rotating Universe \cite{Godel49}
though the first example of a rotating space-time with CTCs was most probably 
found by Lanczos \cite{Lanczos24}. In these cases there is no
universal time because the classical space-time manifold contains
closed time-like curves. However in the following we will show that this strange
behavior can also be viewed as an example of the failure of
classical general relativity on macroscopic length scales.

The hydrodynamic equations for a superfluid that one derives directly from the nonlinear
Schr\"{o}dinger equation are not exactly the classical Euler equations, 
but there are quantum corrections to these equations which become important 
when a certain quantum coherence length becomes comparable to length scale over which the superfluid
density varies. One circumstance where this happens is near the core of a quantized vortex in a rotating superfluid. 

In order to generalize the condensate models of refs. \cite{M96,M97a,M97b,M97c,M98,M07,CM04,M01,M04,CHLS00,CHLS03,RBL03} 
to the case of rotating space-times, we consider the non-linear Schr\"{o}dinger equation in a general stationary space-time 
background described by the line element

\begin{equation}
ds^2 = g_{\mu\nu}dx^{\mu}dx^{\nu} = c_{s}^{2}g_{00}dt^2 + 2c_{s}g_{0i}dtdx^i + g_{ij}dx^idx^j \ , \label{eq1} 
\end{equation}
\noindent where $g_{\mu\nu}$ is time independent. The phenomenological Lagrangian density describing the
condensate of nonrelativistic particles with mass $M$ has the form

\begin{eqnarray}
\mathcal{L} = \sqrt{-g}[\frac{i\hbar}{2}g^{00}\left(\Psi^{\ast}\partial_t\Psi - \Psi\partial_t\Psi^{\ast}\right) +
\frac{{\hbar}^2}{2M}g^{ij}\partial_i\Psi^{\ast}\partial_j\Psi \nonumber\\  +  
\frac{i\hbar c_{s}}{2}g^{i0}\left(\Psi^{\ast}\partial_i\Psi - \Psi\partial_i\Psi^{\ast}\right) +
\frac{\hbar}{2Mc_{s}}g^{0i}\left(\partial_t\Psi^{\ast}\partial_{i}\Psi
+ \partial_t\Psi\partial_{i}\Psi^{\ast}\right) \nonumber\\  +
\mu\Psi^{\ast}\Psi - U(|\Psi|^2)]  \ , \label{eq2}
\end{eqnarray}
\noindent where $g^{\mu\nu}$ is the contravariant tensor inverse
to the metric $g_{\mu\nu}$ for the background space-time, $\mu$ is
the chemical potential, $U(|\Psi|^2)$ is the interaction potential
energy, and $c_{s}$ is the velocity of sound in the condensate at
the equilibrium state. The velocity of sound $c_{s}$ is related to
the interaction potential $U$ by the relations 
$M{c^2_s} = |\Psi|^2U^{''}(|\Psi|^2)$ and $U'(|\Psi|^2) = \mu$. 
From now on we shall set $c_{s}=c$. 
The equation of motion for the condensate order parameter $\Psi$ is

\begin{eqnarray}
\fl i\hbar g^{00}(\partial_t +
{\frac{cg^{0i}}{g^{00}}\partial_i)\Psi} = {\frac{\hbar^2}{2M}}
\frac{1}{\sqrt{-g}}\partial_i({\sqrt{-g}}g^{ij}\partial_j\Psi) + (U^{'} - \mu)\Psi -
{\frac{\hbar}{Mc}}g^{0i}\partial_i\partial_t\Psi \ , \label{eq3}
\end{eqnarray}

\noindent where $g$ is the determinant of the metric.

\vspace{.2cm} It will be useful to write the metric in the form

\begin{equation}
g_{\mu\nu} = \eta_{\mu\nu} + h_{\mu\nu} \ , 
\label{eq5}
\end{equation}

\begin{eqnarray}
ds^{2}=g_{\mu\nu}dx^{\mu}dx^{\nu}=(c dt - A_{i}dx^{i})^{2} - \delta_{ij}dx^{i}dx^{j} \ ,  
\label{eq5}
\end{eqnarray}

\noindent where $\eta_{\mu\nu} = diag(c^{2}, -1, -1, -1)$. 

To first order in $h_{\mu\nu}$ the effect of the background space-time is to introduce 
in the Lagrangian a term $-\frac{1}{2}h_{\mu}^{\nu}T^{\mu}_{\nu}$, where $T^{\mu}_{\nu}$ 
are the four conserved N\"{o}ther currents corresponding to the space and time translational invariance 
of the homogeneous condensate described by (2), and $T_{\mu\nu}$ 
is the symmetrized stress-energy-momentum tensor for the condensate. 
Writing $\Psi = \sqrt{n}e^{iS}$, where $n = |\Psi|^2$ is the number density of particles 
in the condensate, we obtain the velocity field $v_i = \frac{\hbar}{M} \partial_iS$ for the condensate flow. 
This representation of $\Psi$ leads to the steady state quantum hydrodynamic equations for $n$ and $v_i$.

\begin{equation}
\partial_i[n(v_i(1 - \frac{h_{00}}{2c^2}) - h_{ij}v_j -h_{0i})] =
0 \ , \label{eq6}
\end{equation}

\begin{eqnarray}
\frac{\hbar^2}{M\sqrt{n}}{\bnabla}^2\sqrt{n} -
{\frac{\hbar^2}{M\sqrt{n}}}\partial_i(h_{ij}\partial_j\sqrt{n}) +
2(1 - \frac{h_{00}}{2c^2})(\mu - U^{'}) \nonumber\\ - M(1 - \frac{h_{00}}{2c^2}){\bi{v}}^2 + 2Mh_{0i}v_i
- h_{ii}nU^{''} + Mh_{ij}v_iv_j \nonumber\\ -
\frac{\hbar^2}{4M}\nabla^2h_{ii} - \frac{\hbar^2}{2Mc^2\sqrt{n}}\bnabla\cdot(h_{00}\bnabla\sqrt{n})
= 0 \ . \label{eq7}
\end{eqnarray}

It is implicit in our paper that the dynamical role of the nonrelativistic condensate 
is to generate the effective gravitational dynamics a la induced gravity of A. D. Sakharov \cite{ADS67}. 
The difference between Sakharov's induced gravity and the present proposal is as follows. 
Sakharov had envisaged gravity as a result of vacuum polarization due to the presence 
of all relativistic matter fields (`elementary' as well as `composite' particles were included). 
In the condensate model of gravitation the role of the massive nonrelativistic bosons 
is to generate dynamical gravitation (an emergent graviton) and make the computation 
of Newton's gravitational constant $G$ possible. 

The Sakharov idea of induced gravitation \cite{ADS67} 
was incorporated in the Adler program \cite{A82} which turned out to be not too successful for the following reason:  
There is no way one can compute the cosmological constant $\Lambda$ and the gravitational Newton constant $G$ 
in the framework of the relativistic renormalizable quantum field theory (QFT). 
The sign of the Newton constant $G$ is not protected in the Adler type of computation 
(the subtracted dispersion relations) while in the nonrelativistic condensate model 
the sign of the Newton constant is always positive \cite{M88,M92,M07,POM03}. 
One of the responses to the Adler realization of the Sakharov program of induced gravity 
in the context of renormalizable QFT was the so-called no go theorem of Weinberg and Witten (WW) \cite{WW}. 
It is exactly because the condensate models are nonrelativistic that the Newton constant $G$ is computable 
in terms of the mass $M$ of a heavy boson. This also offers the way to elude the WW no-go theorem 
which assumes an exact Poincare invariance.   
However, this effective gravitational dynamics is not generally covariant although it includes the Hilbert-Einstein action 
as the leading term in the expansion in the powers of the quantum coherence length $\xi$ of the condensate. 
The non generally covariant terms in the induced gravitational dynamics are subleading. 
This is the price one has to pay for being able to compute the Newton constant. 
We shall then not concentrate our attention on dissipative modes in the condensate 
but instead we shall study the connection between the superfluid and metric hydrodynamical variables. 
The dissipative modes referred to above are phonons on the superfluid side of the connection and gravitons 
on the gravitational side of the connection between superfluidity and gravitation 
\cite{M88,M92,M96,M97a,M97b,M97c,M98,M07,CM04}. 

In this connection we shall point out that this connection is more subtle than thought before. 
Specifically, we demonstrate on an example of the superfluid quantized vortex that the usual association 
of the velocity field of a fluid and the `acoustic metric' does not describe the physics correctly. 
As it happens, it is just because the map between fluid variables and the so-called acoustic metric 
was made the basis of all analogue models \emph{for} gravitation, our simple result presented 
in what follows invalidates all this broad area of analogue gravity models. 
In other words, the `acoustic metric' has nothing to do with natural phenomena, 
and, in particular, with gravitation and cosmology.  

In the following we will find the classical metrics which correspond to superfluid flows with vortices
when $h_{00} = 0$ and $\partial_{_3}g_{\mu\nu} = 0$. The metric in
our action is not a dynamical field. Instead the metric components
only act as Lagrange multipliers. The role of these Lagrange
multipliers is to enforce the local equilibrium in the condensate.
The homogenous vacuum state of the condensate is characterized by
$|\Psi| = const$, $g_{\mu\nu} = \eta_{\mu\nu}$ and $U^{'} = \mu$.

\vspace{.2cm} We first seek a solution of (6) and (7)
corresponding to a single vortex in the condensate. The phase $S$
of the condensate corresponding to a single vortex has a simple
form: $S = N\varphi$, where $\varphi$ is the azimuthal angle
defined by the formula $\varphi = Arctan(\frac{x^2}{x^1})$ and $N$
is the vortex number which is an integer. The velocity field
corresponding to the vortex configuration is:

\begin{equation}
v_i = N\frac{\hbar}{M} \partial_i\varphi= - {N\kappa\over {2\pi r^2}}\epsilon_{ij}x_j  \ , \label{eq8}
\end{equation} 
\noindent where $\kappa = \frac{h}{M}$ is the fundamental unit of
quantized circulation $\oint \bi{v}\cdot
d\bi{l}$ or the flux of the vorticity field
$\omega_{ij} =
\partial_iv_j - \partial_jv_i$. 

The velocity field of a vortex $v_i$ has the form of the Aharonov-Bohm
electromagnetic potential \cite{AB59} while the vorticity $\omega
= \frac{1}{2}\epsilon_{ij}\omega_{ij} =
\epsilon_{ij}\partial_iv_j$ is an analog of the Aharonov-Bohm
magnetic field produced by an infinitely thin solenoid, $\omega =
\kappa\delta(x^1)\delta(x^2)$.

\vspace{.2cm} It turns out that because of the presence of the
potentials $h_{0i}$ and $h_{ij}$ in the hydrodynamic equations the
superfluid density $n$ will be nearly constant when $r$ is greater
than the coherence length $\xi = \frac{\hbar}{Mc}$. Indeed it
is straightforward to show that if $n$ is constant and the
velocity has the form given in (8), then (6) and (7) have
a solution for $N = 1$

\begin{equation}
h_{00} = 0   \qquad    h_{0i} = -c^{-1}v_i   \qquad  h_{ij} =  c^{-2}{v_iv_j} \ , \label{eq9}
\end{equation}
\noindent where $c=c_{s}$, of course. These values for the potentials $h_{\mu\nu}$ are
equivalent to the metric for the background space-time of the
local `spinning cosmic string' solution of the Einstein field
equations \cite{M95,M86,Star63} in the region where $n$ is constant,
i. e. for $r \gtrsim \xi$. The line element for this solution (for
$r
> 0$) has the form \cite{M95,M86}

\begin{equation}
ds^2 = (c dt - Ad\varphi)^2 - dr^2 - r^2d{\varphi}^2 - dz^2  \ , \label{eq10}
\end{equation}
\noindent where $A = {\kappa\over {2\pi c}} = \xi$. The
string-like singularity at $r=0$ has neither mass density nor
pressure, so space-time is flat for $r > 0$. However, the string
rotates resulting in frame dragging. This frame dragging is
represented by the appearance of a vector potential $A_i$
\cite{M95,M86} with azimuthal component $A_{\varphi} = A =
{\kappa\over {2\pi c}}$. The frame dragging implied by the
metric (10) is evidently closely related to the velocity field
surrounding a single vortex filament in a superfluid. Indeed De
Witt pointed out some time ago \cite{Bryce66} that the vector
potential $A_i$ associated with frame dragging can be formally
identified as the vector potential for a superconductor. Kirzhnits
and Yudin \cite{Kirz95} have also studied stationary superfluid
flows in the presence of gravitational fields $g_{0i}$ produced by
rotating compact, massive objects (superfluid cores of neutron
stars). Balasin and Israel \cite{BI99} have concluded
that vortex filaments in a superfluid neutron star do produce
gravimagnetic forces, contrary to the statements in the
literature.

\vspace{.2cm} It should be noted that collective bosonic excitations 
in the condensate and other massless or massive fermionic and bosonic excitations (impurities) 
will feel the gravitational field (10) associated with the gravitational vortex. 
However, this gravitational field is not the same as the acoustic metric \cite{Vol98,FV03}. 
The scattering cross-section for fermionic (bosonic) particles will be
given by the Aharonov-Bohm cross-section \cite{M95,M86} as is
the scattering of quasiparticle excitations of unit electric
charge on Abrikosov vortices \cite{Ab57} in type II
superconductors. In this sense the `spinning cosmic string' is a
gravitational analog of the Abrikosov vortex \cite{M95,M86}. 
This is also the reason why one of the authors has called the
scattering of relativistic particles by gravitational vortices the
gravitational Aharonov-Bohm effect \cite{M95,M86}. The
scattering cross-section for quasiparticles propagating in an acoustic metric associated 
with a vortex in superfluid He3 has been given in ref. \cite{Vol98} 
and for the reasons just mentioned \emph{is not the same} as the gravitational Aharonov-Bohm scattering
cross-section \cite{M95,M86}. 
The present paper demonstrates clearly, following \cite{M95,M86}, 
that in the superfluid model of an emergent gravitation 
the `spinning cosmic string' metric describes a gravitational vortex \cite{M86,CM04} 
and the Aharonov-Bohm scattering cross-section \cite{M95,CM04,AB59,M86} 
describes interaction of fermionic quasiparticles and bosonic collective excitations with this vortex.

\vspace{.2cm} The space-time corresponding to the metric (10) does
not have a universal time because closed time-like curves appear
close to the axis of the gravitational vortex. What does not seem
to have been noted before, though, is the fact that closed
time-like curves appear in the gravitational vortex background
(10) at exactly the radius where a classical hydrodynamic
description of the superfluid begins to fail. Indeed the
superfluid velocity (8) will become comparable to the velocity of
sound $c$ when the radius $r$ is close to the quantum
coherence length $\xi$. Therefore superfluid rigidity and
classical hydrodynamics break down as one enters the core of the
vortex. Remarkably, this breakdown of a classical description of
the superfluid seems to be closely related to the breakdown of
causality in classical GR associated with the formation of closed
time-like geodesics. The condition for the appearance of closed
time-like curves in a rotating space-time is that
$g_{\varphi\varphi} > 0$, which for the gravitational vortex
metric (10) becomes the condition

\begin{equation}
r < r_{c} = {\kappa\over 2\pi c} = \xi \ . \label{eq11}
\end{equation}
\noindent That is, closed time-like curves appear in the
gravitational vortex solution of the Einstein equations near to
the axis of the string where the velocity of frame dragging
exceeds the speed of light. In the superfluid picture this
corresponds to the core of the vortex where the superfluid flow
velocity exceeds the speed of sound $c$. As previously
discussed, this is just where a classical hydrodynamic description
of the fluid flow in a quantized superfluid vortex breaks down.
Indeed, the solution to the equations of quantum hydrodynamics in
the presence of the potentials $h_{\mu\nu}$ given by (9) is valid only 
in the region where the condensate particle
density $n$ is constant. The corresponding space-time metric (10)
is perfectly well behaved in this region ($r > \xi$). It is only
after the na\"{\i}ve analytic continuation of the metric (10) to
the region $r < \xi$ is attempted that the causality violating
regions appear in the space-time.

\vspace{.2cm} This observation provokes one to ask if the
appearance of closed time-like curves in solutions of the
classical Einstein field equations might always be associated with
a breakdown of superfluid rigidity? In particular, one might
wonder if the appearance of closed time-like curves in
G\"{o}del-like universes is related to the behavior of rotating
superfluids. The G\"{o}del metric for a rotating universe can be
written in the form \cite{Reb83}

\begin{equation}
ds^2 = (cdt +  \Omega(r)d\varphi)^2 - dr^2 - f^2d\varphi^2 - dz^2
\ , \label{eq12}
\end{equation}

\noindent where 
\begin{equation}
\Omega(r) =
\frac{4\Omega}{m^2}sinh^2(\frac{mr}{2})   \qquad   f(r) = \frac{1}{m}sinh(mr)  \  . \label{eq13}
\end{equation}
\noindent  
In the limit of small $r$, $\Omega(r)$
approaches $\Omega r^2$, i.e. the off-diagonal metric component
$g_{0\varphi}$ equals the velocity potential inside a body rigidly
rotating with angular velocity $\Omega$. It can be seen that the
metric component $g_{0\varphi}$ for the G\"{o}del universe has a
very different dependence on radius from that of the gravitational
vortex; however, as we shall now see this very different behavior
is characteristic of what happens in a rapidly rotating
superfluid.

\vspace{.2cm} Feynman pointed out \cite{Feyn55} that when many
vortices are present the velocity field in the superfluid averaged over the distance scales 
which are large comparing to the inter-vortex distances will approach the velocity field of a rigidly rotating body, 
i.e. $\bi{v} = \bi{\Omega}\times\bi{r}$. The true superfluid flow must be locally curl-free, 
and hence one has to introduce the `phenomenological device' of a back flow \cite{Feyn55,Feyn72}. 
When the area density $\sigma$ of
vortices is not too high it is reasonable to approximate the phase
in (8) as a sum $S = \sum_a Arg(w - w_a)$, $w = x^1 + ix^2$,
of phases of individual vortices each with vortex number $N = 1$. 
The velocity field in this approximation can be
written in the form

\begin{equation}
v_i = \frac{\kappa}{2\pi}\partial_iS = -
\frac{\kappa}{2\pi}\epsilon_{ij}\partial_j\sum_aln|\bi{x}
- \bi{x}_a| \ . \label{eq14}
\end{equation}
\noindent Evaluating the vorticity $\omega =
\epsilon_{ij}\partial_iv_j$ and replacing the sum in (14) by
an integral we obtain
\begin{equation}
\omega = \frac{\kappa\sigma}{2\pi}\nabla^2_x\int d^2y
ln|\bi{x} - \bi{y}| \ . \label{eq15}
\end{equation}
\noindent Using the relation 

\begin{equation}
\nabla^2_xln|\bi{x} -
\bi{y}| = {2\pi}\delta^{(2)}(\bi{x} -
\bi{y})  \  , \label{eq16}
\end{equation}
\noindent 
we obtain $\omega = \kappa\sigma$. It follows then that

\begin{equation}
v_i = -
\frac{\kappa\sigma}{2}\epsilon_{ij}x_j  \  . \label{eq17}
\end{equation}
\noindent This means that coarse-grained 
velocity field of a lattice of vortices is indeed that of a rigid body \cite{Feyn72} rotating with the
angular velocity $\Omega = \frac{\kappa\sigma}{2}$.

\vspace{.2cm} Since the gravitational vortex solution (10) is
spatially flat, it makes sense to construct a new solution to the
Einstein equations by simply superposing the velocity fields 
(14) corresponding to a collection of parallel gravitational
vortices. Following the same line of reasoning that leads one to
rigid body rotation in the case of many superfluid vortices, one
would surmise, based on the identification $h_{0i} = -c^{{-1}}v_i$, that in
the presence of many gravitational vortices the metric of
space-time would assume the form
\begin{equation}
ds^2 = (c dt +  \frac{1}{c}\Omega r^2 d\varphi)^2 - dr^2 -
r^2d\varphi^2 - dz^2
  \ . \label{eq18}
\end{equation}
\noindent The metric constructed in this way produces via the Einstein equations 
an effective stress-energy tensor which, of course, corresponds to unphysical sources.  
In the same way the rigid body rotation velocity field is produced by the coarse-grained 
locally smoothed out curl-free velocity field due to a distribution of vortices, and  
in the same way one produces an effective smooth distribution of mass described by a mass density 
$\rho(\bi{x},t)$ out of the discrete distribution of point-like masses. Of course, this procedure produces mass 
in an empty space between point-like massive constituents of matter. 
This should be easily understood as an application of the basic principles of hydrodynamics. 
Everybody knows that `normal' hydrodynamics fails on the distance scale of coarse-graining, this is to say 
on the inter-atomic or inter-molecular distances. 
The same is true for our prescription of producing stationary rotating space-times from the distribution 
of localized quantized vorticity in the superfluid substratum underlying the metric continuum $g_{\mu\nu}(x)$.  
 
The metric (18) is in fact just the non-vacuum Som-Raychaudhuri
solution of the Einstein field equations
\cite{Reb83,SomRay68}. This metric can be obtained
from the G\"{o}del metric (12), (13) by letting $m\rightarrow 0$. It
can be seen that the velocity of frame dragging for the metric
(18) is just the velocity inside a rigidly rotating body. The
condition for the appearance of closed time-like curves, i. e.
$g_{\varphi\varphi} > 0$, in the Som-Raychaudhuri space-time is

\begin{equation}
\Omega r_{c} > c \ . \label{eq19}
\end{equation}
\noindent That is, closed time-like curves appear when the
velocity of frame dragging exceeds the speed of light. In contrast
with the gravitational vortex closed time-like curves appear in
the Som-Raychaudhuri space-time at large radii. The appearance of
closed time-like curves in G\"{o}del space-times mimics the
behavior of Som-Raychaudhuri space-time in that the closed
time-like curves appear at large radii. In particular, for the
G\"{o}del metric (12), (13) the condition for the appearance of closed
time-like curves is

\begin{equation}
\frac{2\Omega}{m}tanh\frac{mr_c}{2} > c  \ . \label{eq20}
\end{equation}

\noindent When $m\rightarrow 0$ this condition reduces to 
(19). When $m = {2\Omega\over c}$ the radius where the velocity of
frame dragging approaches the speed of light recedes to infinity,
and the space-time will be free of closed time-like curves
everywhere. We now wish to inquire as to the significance of the
conditions (19) and (20) from the point of view of a rotating
superfluid. Evidently then a superfluid description for the
metrics (12), (13), and (18) will require an external rotating container
of normal matter to create a frame dragging potential. 
One may not analytically continue the metric beyond the surface of a container 
which rotates with the speed of sound in a superfluid. 
The walls of a container are a part of a different from the superfluid substratum phase of matter.  

The occurrence of solid body-like frame dragging in the G\"{o}del and
Som-Raychaudhuri metrics may seem to be incompatible with a
superfluid interpretation for space-time because
$\bi{\nabla}\times\bi{v} = 2\bi{\Omega}$ for a solid body
rotating with angular velocity $\bi{\Omega}$, whereas the flow
velocity of a superfluid must have zero curl since it is the
gradient of a phase. The resolution of this paradox is that the
solid body rotation curve corresponds to a coarse-grained average
of the velocities from an array of individual vortices. 
In between the vortices the flow is irrotational so
$\bi{\nabla}\times\bi{v} = 0$ in the superfluid condensate. 
The phenomenological concept of a back-flow should enter the discussion 
here \cite{Feyn55,Feyn72} but we shall not dwell on it here. 
The metric field $g_{\mu\nu}$ is, of course, a coarse-grained hydrodynamical variable. 

\vspace{.2cm} In contrast with the case of a single vortex, the
coarse-grained potentials associated with the array of vortices do
not satisfy the time independent hydrodynamic equations (6) and (7).
Indeed, in contrast with the case of the single vortex, the term
${\bnabla}^2 h_{ii}$ in (7) which comes from the quantum pressure
no longer cancels the term $h_{ij} v_i v_j$ which arises as a
relativistic correction to the kinetic energy density of the
condensate. Although a simple superposition (14) of the
single-quantized vortex solution (8), does not satisfy the
superfluid equations (6) and (7), there do exist multi-vortex solutions. In
particular, there exist time independent solutions representing a
regular lattice of vortices, the Tkachenko lattice
\cite{Tkach65}.

\vspace{.2cm} When an impulse of energy is applied to a very low
temperature rotating superfluid condensate, then a turbulent state
containing a time dependent tangle of quantum vortices can develop
\cite{Vin02}. Such a regime is known as quantum turbulence. If
space-time is indeed a condensate and the conditions for the
development of quantum turbulence, i. e. rotation and an impulse
of energy, are met, then there should be characteristic
observational signatures. For example, the onset of quantum
turbulence in cosmological space-times would lead to a
characteristic scale-free spectrum of energy density fluctuations. 
Indeed, at the largest distance scale the spectrum of quantum turbulence is the same as the celebrated Harrison, Peebles-Yu, 
and Zel'dovich \cite{h70,py70,z72} spectrum of primordial energy density fluctuations \cite{POM03}. 

\vspace{.2cm}
Few remarks on the nonrelativistic character of our superfluid model of stationary and rotating spacetimes are in order. 
This model should be regarded as a model of an `emergent gravitation'. 
The idea of the  `atomic model of gravitation', or `the constituent model of gravitation' 
\cite{M95,M96,M97a,M97b,M97c,M98,M07,CM04}, which essence is that the problem of gravitation 
should be regarded as an example of a quantum many body problem 
was first proposed in refs. \cite{M92,M95,M96,M97a,M97b,M97c,M98,M07,CM04}. 
The foundations of the microscopic theory of gravitation should be based on the idea of the existence 
of fundamental (bosonic) constituents \cite{M92,M95,M96,M97a,M97b,M97c,M98,M07,CM04} 
and the emergence of macroscopic quantum states \cite{M92,M96,M97b,M97c,M98,M07} characterized by the presence 
of the off-diagonal long range order \cite{GC92}. 
This approach to gravitation has later led to the proposed solution of the problem 
of the final state of gravitational collapse of quantum matter and to the work on emergent relativity 
in a BEC \cite{M07,CM04,M01,M04,CHLS00,CHLS03,RBL03}. 
The quantum many body models considered in the context of emerging relativity and gravitation 
are by construction nonrelativistic. In view of what was said above it would be an example 
of a circular logic if one considered relativistic models of Bose-Einstein condensates 
in the context of the problem of gravitation. For these reasons the fully relativistic model 
of a BEC coupled to a dynamical gravitational field is beyond the scope of this paper. 

In order to cleanly separate the issue of an `emerging graviton' from the nonperturbative aspects of gravitation, 
such as the presence of gravitational vortices, we have focused on non-riadiating field configurations on two sides 
of the relationship between bosonic superfluid variables and gravitational field variables. 
We have considered stationary axisymmetric field configurations of a nonrelativistic BEC coupled 
to a \emph{nondynamical} gravitational field. In fact, we have demonstrated in this paper that the relationship between 
the superfluid variables and the stationary gravitational field variables does exist. 

It should be clearly recognized that this direction of thinking about the microscopic nature of gravitation 
is completely orthogonal to other points of view on the quantum nature of gravitation which were in evidence 
in the published literature prior to \cite{M92,M95,M96,M97a,M97b,M97c,M98,GC92,CM04} 
when the idea of `constituent model of gravitation' has emerged and later around 2000-2001 
when the idea of `gravastars' and `dark energy stars' was first published \cite{M07,CM04,M01,M04,CHLS00,CHLS03,RBL03}. 
For example this way of thinking about gravitation is completely orthogonal to the ideology 
of string models, `loop' gravity, and analogue models.    
In particular it should be clear that the so-called analogue models \emph{for} but not \emph{of} 
gravitation were invented to mimic the geometric aspects of the Einsteinian description of gravitation. 
This ideology found its expression in the concept of the `acoustic metric'. 

The `acoustic metric' of analogue models has nothing to do with gravitation and the fact that it is not dynamical 
is not relevant to this conclusion. It is simply a wrong map between fluid field variables and the metric variables. 
The map clearly does not distinguish between ordinary fluids and superfluids.
In order to illustrate this point it is sufficient to examine the `acoustic metric' corresponding 
to the Onsager-Feynman vortex in a superfluid.  One finds that it corresponds to a highly curved \cite{FV03}
rather than a \emph{locally flat} spacetime \cite{M86,CM04}.  

It is clear that `apples do not fall on a ground' in the `worlds' of analogue models 
and string models have nothing to do with gravitation in $3+1$ dimensions. 
Saying that something is analogous to something else and that graviton is built in into string models 
is not the same as establishing the fundamental physical principles which lead to the microscopic explanation 
of gravitational phenomena. These Principles are Quantum Theory and Atomism. 
By Atomism we mean the Hypothesis which posits the Existence of Massive Constituents.  
Then Quantum Coherence, Quantum Entanglement, and the Emergence of Macroscopic Quantum States of New Kind(s) 
of Matter must lead to Gravitational Phenomena. The problem of microscopic explanation of gravitational phenomena 
is then reduced to the search for the 3D universality class of quantum critical phenomena. The hint is coming here 
from the proposed solution of the `black hole event horizon' problem \cite{M01,M04,CHLS00,CHLS03,RBL03}. 
   
\ack The authors would like to thank their colleagues James D. Bjorken and Robert B. Laughlin 
for valuable comments on the subject of this paper. 
We also wish to thank the referees for their insightful remarks. 
One of us (G. C.) would like to acknowledge hospitality extended to him 
at the University of South Carolina. This material is based upon work
(partially) supported by the National Science Foundation under
Grant No. 0140377 (P. O. M.). This work was also performed (in
part) under the auspices of the U.S. Department of Energy by
University of California Lawrence Livermore National Laboratory
under contract No. W-7405-Eng-48 (G. C).

\end{document}